\documentclass[pra,final,twocolumn,showpacs,superscriptaddress]{revtex4-1}
\usepackage{graphicx,amsmath,amsfonts,wasysym,color}

\begin{document}

\title{Rapid crystallization of externally produced ions in a Penning trap}

\author{T. Murb\"ock}
\email{these authors have contributed equally to the presented work}
\affiliation{Institut f\"ur Angewandte Physik, Technische Universit\"at Darmstadt, 64289 Darmstadt, Germany}
\author{S. Schmidt}
\email{these authors have contributed equally to the presented work}
\affiliation{Institut f\"ur Kernphysik, Technische Universit\"at Darmstadt, 64289 Darmstadt, Germany}
\affiliation{Institut f\"ur Kernchemie, Johannes Gutenberg-Universit\"at Mainz, 55099 Mainz, Germany}
\author{G. Birkl}
\affiliation{Institut f\"ur Angewandte Physik, Technische Universit\"at Darmstadt, 64289 Darmstadt, Germany}
\author{W. N\"ortersh\"auser}
\affiliation{Institut f\"ur Kernphysik, Technische Universit\"at Darmstadt, 64289 Darmstadt, Germany}
\author{R.C. Thompson}
\affiliation{Department of Physics, Imperial College London, London SW7 2AZ, UK}
\author{M. Vogel}
\affiliation{GSI Helmholtzzentrum f\"ur Schwerionenforschung, 64291 Darmstadt, Germany}

\begin{abstract}
We have studied the cooling dynamics, formation process and geometric structure of mesoscopic crystals of externally produced magnesium ions in a Penning trap. We present a cooling model and measurements for a combination of buffer gas cooling and laser cooling which has been found to reduce the ion kinetic energy by eight orders of magnitude from several hundreds of eV to $\mu$eV and below within seconds. With ion numbers of the order of $10^3$ to $10^5$, such cooling leads to the formation of ion Coulomb crystals which display a characteristic shell structure in agreement with theory of non-neutral plasmas. We show the production and characterization of two-species ion crystals as a means of sympathetic cooling of ions lacking a suitable laser-cooling transition.
\end{abstract}

\maketitle

\section{Introduction}
Laser cooling is an effective tool to reduce the temperature of confined ions, particularly from temperatures of up to several thousands of kelvin down to the Doppler limit, which is commonly in the mK range \cite{ita,buch,dem,mett}.
For magnesium ions, this has been demonstrated under various confinement conditions \cite{nag,rct,died,bir,dho}. 
Such cooling is beneficial for the stable confinement in traps over extended periods of time \cite{ita,torr}, and essential for precision spectroscopy as it reduces spectral line broadening caused by the Doppler effect \cite{mett,dem}. For other systems, including highly charged ions, laser cooling is not a method of choice, owing to the lack of suitable (fast) optical transitions \cite{pr}. Resistive cooling \cite{ita} can be an effective method for such systems, especially if they carry high electric charge. However, the minimal energy is usually limited to energies which correspond to the ambient temperature on the scale of several kelvin \cite{rcool}. Hence, sympathetic cooling with simultaneously confined laser-cooled ions is a good possibility for these ions to reach the mK regime \cite{ita,piet}. 

Here, we discuss laser cooling of singly charged magnesium ions in a Penning trap \cite{gab89,werth,gho}, following their dynamic capture \cite{schn,spec1} from an external source. In such situations, they commonly have high initial energies unsuitable for efficient laser cooling. Under conditions similar to the present ones, laser cooling times have been reported to be of the order of many minutes \cite{gru}. We have found that a combination of laser cooling and buffer gas cooling is capable of reducing the ion kinetic energy by more than 8 orders of magnitude within seconds. The ions `crystallize' into structures given by their mutual Coulomb repulsion in the presence of the confining trap potential, similar to the results in \cite{died,bir,dre,dre2,horn,mit,rich}, for which we find agreement with non-neutral plasma theory. The mesoscopic size of several thousands to several tens of thousands of ions is advantageous for sympathetic cooling, as such crystals are large enough to provide a sufficiently large cold bath for other charged particles to be cooled. 

\section{Experimental Setup}
\label{setsec}
\begin{figure}[h!]
\begin{center}
 \includegraphics[width=0.8\columnwidth]{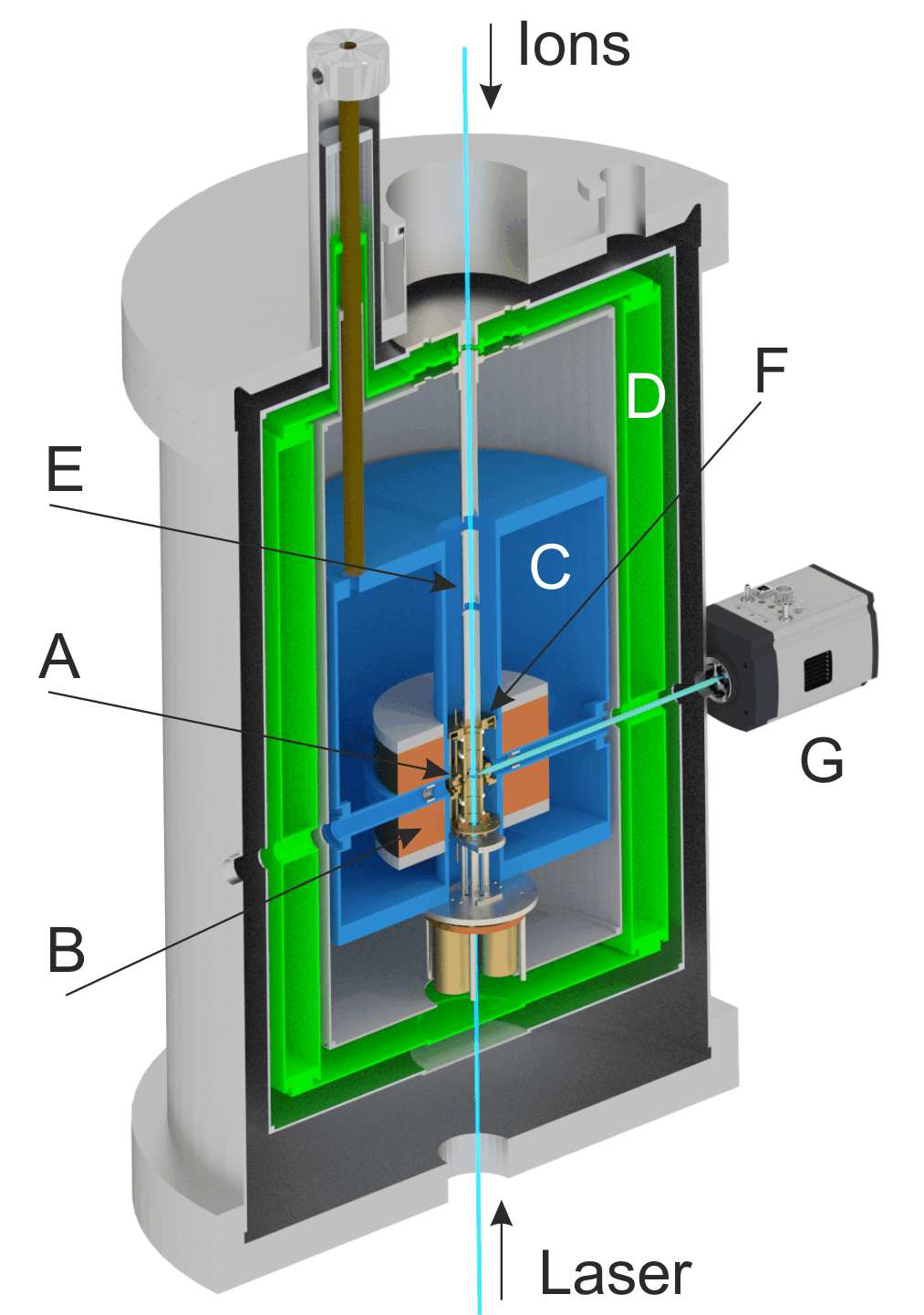}
  \caption{\small Sectional view of the SpecTrap setup. (A) Penning trap, (B) Magnet, (C) LHe dewar, (D) LN$_2$ dewar, (E) pulsed drift tubes, (F) non-destructive ion detector, (G) CCD camera. For details see text.}
  \label{set}
\end{center}
\end{figure}

The experiments have been performed with the SpecTrap experiment \cite{spec0,spec1} located at the HITRAP facility \cite{kluge} at GSI and FAIR, Germany.
The experimental setup (Fig. \ref{set}) 
has previously been described in detail in \cite{spec1}. 
Briefly, a cylindrical Penning trap is located in the homogeneous field of a superconducting magnet and is cooled to liquid-helium temperature.
Fig. \ref{set} shows a sectional view of the setup with the Penning trap (A) installed in the cold bore and in the center of the magnetic field of the surrounding superconducting magnet (B) of Helmholtz geometry \cite{helm}. The cold bore with the trap and its cryo-electronics is cooled by liquid helium (C) which is shielded by liquid nitrogen (D). 

The ions are transported into the trap from above, via a low-energy UHV beamline connecting the ion sources with the trap. Ions can be obtained either from a dedicated pulsed source of singly charged ions \cite{sou}, from other external ion sources such as electron beam ion sources \cite{sparc}, or from the HITRAP low-energy beamline \cite{bea}.
\begin{figure}[h!]
\begin{center}
 \includegraphics[width=0.75\columnwidth]{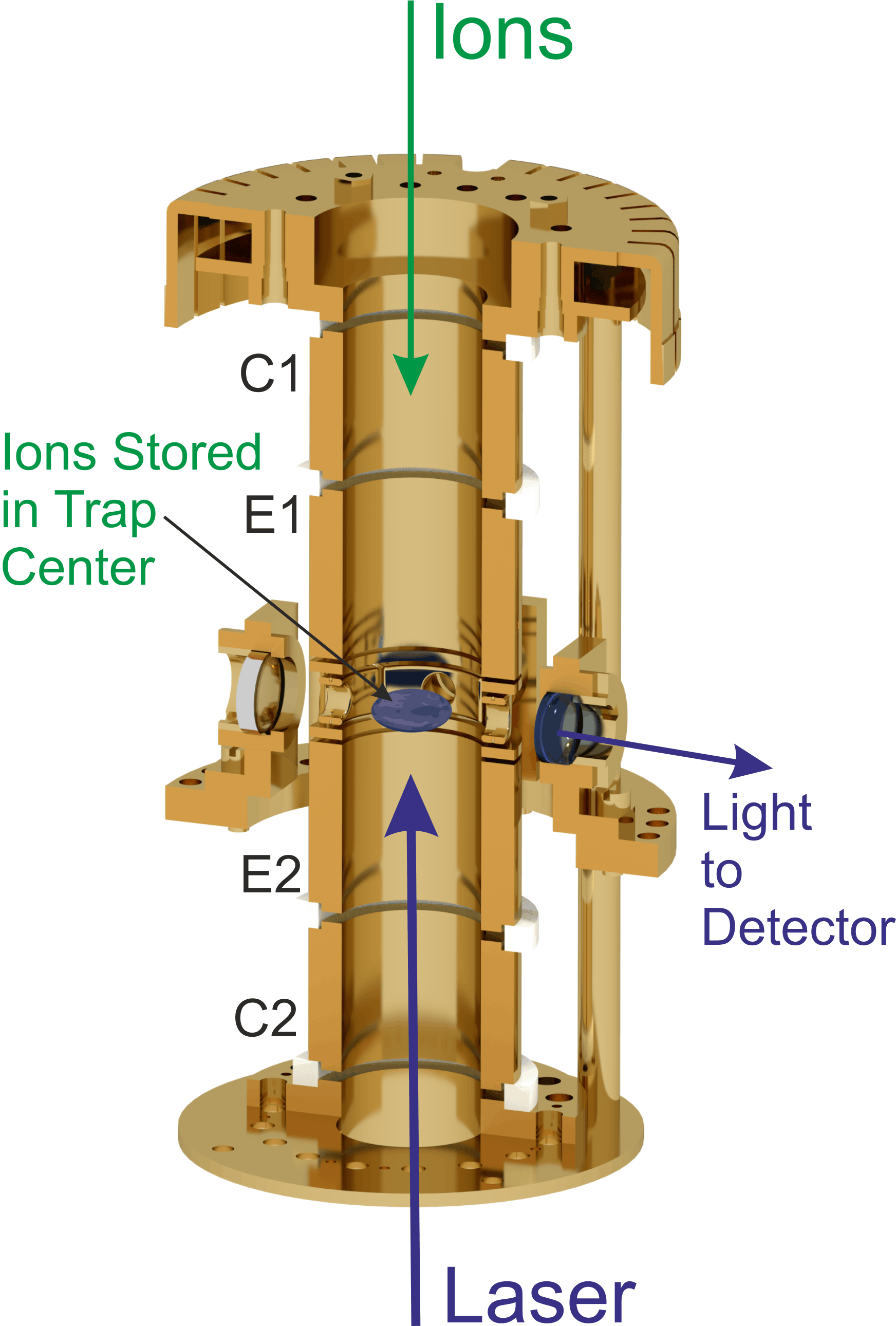}
  \caption{\small Schematic of the SpecTrap Penning trap.}
  \label{trap}
\end{center}
\end{figure}
A set of pulsed drift tubes (E) \cite{book,keefe} located above the trap allows the deceleration of ion bunches from transport energies of the order of keV per charge to energies of the order of 100\,eV per charge, suitable for dynamic capture \cite{schn} and subsequent storage in the trap center. 

The ion number can be estimated from a non-destructive measurement of the induced charge signal when the ion bunch enters the trap. To that end, a dedicated low-noise charge amplifier detector (F) has been built and operated \cite{nid}. Imaging of the stored ions is done via radial ports with an outside CCD camera (G).

Fig. \ref{trap} shows a sectional view of the Penning trap. It is a cylindrical open-endcap 5-pole Penning trap \cite{gab89} (one segmented ring and two compensation electrodes between endcap electrodes E1 and E2) with additional capture electrodes C1 and C2. The latter are used for dynamic capture \cite{schn,spec1} of externally produced ions by creating a potential well after incoming ions have entered the trap. 

The ring electrode located in the optical plane is split into four segments for radial electronic excitation and detection, with one hole of 4.8\,mm diameter in each segment for optical access. Radial ports in the plane of the trap center guide the fluorescence light via a two-lens system to the outside photon counter and CCD camera (EM-CCD C9100-24B, Hamamatsu).
The light collection solid angle amounts to 0.09\,sr or 0.7\% of $4\pi$. A first lens with 25\,mm focal length and located at that
distance from the trap center collimates the light, and a
second lens with 150\,mm focal length focuses it on the detector. The measured magnification factor of the system under the present conditions is 4.69(8). At a detector pixel size of 13\,$\mu$m, this corresponds to a resolution of about 3\,$\mu$m, which by optical imperfections is increased to about 10\,$\mu$m for the present observations.

The excitation laser for optical detection and laser cooling enters the trap from below, along the central axis ($z$-axis).
Laser cooling of the stored $^{24}$Mg$^+$ ions is achieved on the red-detuned side of the 279.55-nm $^{2}\text{S}_{1/2} \rightarrow {^{2}\text{P}_{3/2}}$ transition with a natural linewidth of $\Gamma=2\pi \times 41.8 \times 10^6$\,s$^{-1}$ \cite{nist}.
Frequency-quadrupling of light from a commercial infrared fiber laser produces the required light with a spectral width of less than 1\,MHz and a few tens of mW of maximum power. The laser can be tuned at a rate of up to 200\,MHz/s \cite{caz}.

In short, an experimental cycle consists of the following steps:
\begin{itemize}
\item ion bunch production in an external source 
\item transport at energies of up to 5\,keV per charge
\item ion deceleration in pulsed drift tubes
\item dynamic ion capture in the trap
\item ion cooling and spectroscopy, CCD imaging.
\end{itemize}
Ions can be accumulated (`stacked') by capturing additional ion bunches while ions remain confined in the trap \cite{ros}. Hence, it is possible to subsequently load the trap from different ion sources. It is one advantage of Penning traps that a broad range of different mass-to-charge ratios can be stored simultaneously \cite{werth,gho}.

Singly charged magnesium ions from the pulsed external source have been captured and stored for studies of the temporal dynamics of ion cooling and crystal formation, and their geometric properties. Upon dynamic capture, the ions have kinetic energies of up to several hundreds of eV per charge (typically, 400\,eV per charge have been used), which is far outside the realm of efficient laser cooling. Therefore, a combination of buffer gas and laser cooling is applied. The following section is dedicated to a model of the expected cooling behaviour.

\section{cooling model}
\label{coolmod}
We have developed a simple yet realistic description of the effect of combined buffer gas and laser cooling. It extends the semi-classical model of Doppler cooling as presented in \cite{Wes07} in order to describe the evolution of the fluorescence signal during the formation process of ion Coulomb crystals in a Penning trap. Based on a rate-equation formalism, we find analytic solutions for the time dependence of the energy and thus for the fluorescence rate of a single particle. To this end, we add a recoil-heating term and an exponential cooling term to the rate equation. The latter accounts for the buffer gas cooling \cite{ita}. The laser frequency is tuned linearly with time.  

According to the formalism presented in \cite{Wes07}, the scaled energy $\epsilon$ of a single particle with mass $m$ confined in a harmonic potential has a time derivative given by  
\begin{eqnarray}
\frac{d\epsilon}{d\tau}=&-&\gamma_{1}(\epsilon-\epsilon_{1}) +\frac{4}{3}r\frac{1}{2\sqrt{\epsilon r}}\text{Im}(Z)\nonumber\\&+&\frac{1}{2\sqrt{\epsilon r}}\left(\text{Re}(Z)+\delta\,\text{Im}(Z)\right),
\label{Eq:1}
\end{eqnarray}
with $Z=i/\sqrt{1-(\delta+i)^2/4\epsilon r}$. In this equation, the energy $E$ of the particle, the laser detuning $\Delta$ and the recoil energy $E_{\text{R}}=(\hbar k_{z})^2/2m$ are scaled by the energy $E_0$ such that 
$\{ \epsilon , \delta , r\}  \equiv \{ E , \hbar\Delta , E_{\text{R}} \}/E_{0}$.
Here, $E_0 \equiv \hbar\Gamma\sqrt{(1+s_0)}/2$, where $\Gamma\sqrt{(1+s_0)}$ is the power-broadened linewidth. The on-resonance saturation parameter $s_0$ is determined according to $s_0=I/I_0$, where $I$ is the intensity of the laser at the position of the ions and $I_0$ is the saturation intensity.
The time $t$ is scaled by $t_0$ which is the inverse of the on-resonance fluorescence rate, such that $\tau \equiv t/t_{0}$, where $t_0$ is given by 
\begin{equation}
t_{0}=\left(\frac{\Gamma}{2}\frac{s_0}{1+s_0}\right)^{-1}. 
\end{equation}
$\Gamma$ is the decay rate of the excited state, $k_{z}$ is the $z$-component of the excitation laser wave vector, $\gamma_{1}$ is an exponential cooling rate factor and $\epsilon_{1}$ is the minimum energy that can be achieved by this exponential cooling mechanism. It is represented by the first term in Eq.\,(\ref{Eq:1}), which for buffer gas cooling is given by \cite{mk}
\begin{align}
\gamma_{\text{1}}=\frac{q}{m}\frac{1}{\mu_{0}}\frac{p/p_{0}}{T/T_{0}}.
\label{9}
\end{align}
The damping coefficient $\gamma_1$ depends on the ion mobility $\mu_{0}$ of the buffer gas, the residual gas pressure $p$ and the temperature $T$ of the buffer gas normalized by the standard pressure $p_{0}$ and the standard temperature $T_{0}$, respectively. The second and third terms of Eq.\,(\ref{Eq:1}) describe the laser-particle interaction, including stochastic heating of the particle due to photon recoil and laser Doppler cooling. The factor $4/3$ in the second term is true for isotropic emission characteristics. The scaled fluorescence rate can be found to be \cite{Wes07}
\begin{align}
\gamma_{\text{sc}} \equiv \frac{dN_{\text{ph}}}{d\tau}=\frac{1}{2\sqrt{\epsilon r}} \text{Im}(Z).
\label{Eq:3}
\end{align}
The laser detuning has the form $\Delta=\Delta_{i}+\Delta_{m}\times t$, where $\Delta \equiv \omega-\omega_0$ is the detuning of the actual laser frequency $\omega$ from the resonance frequency $\omega_0$. $\Delta_i$ is the initial laser detuning at time $t=0$ and $\Delta_{m}$ is the scan rate. The time $t=0$ denotes the start of the cooling process; this corresponds to the time when the ions are captured in the trap where buffer gas and cooling laser beam are present. 

\begin{figure}[h]
\centering
\includegraphics[width=\columnwidth]{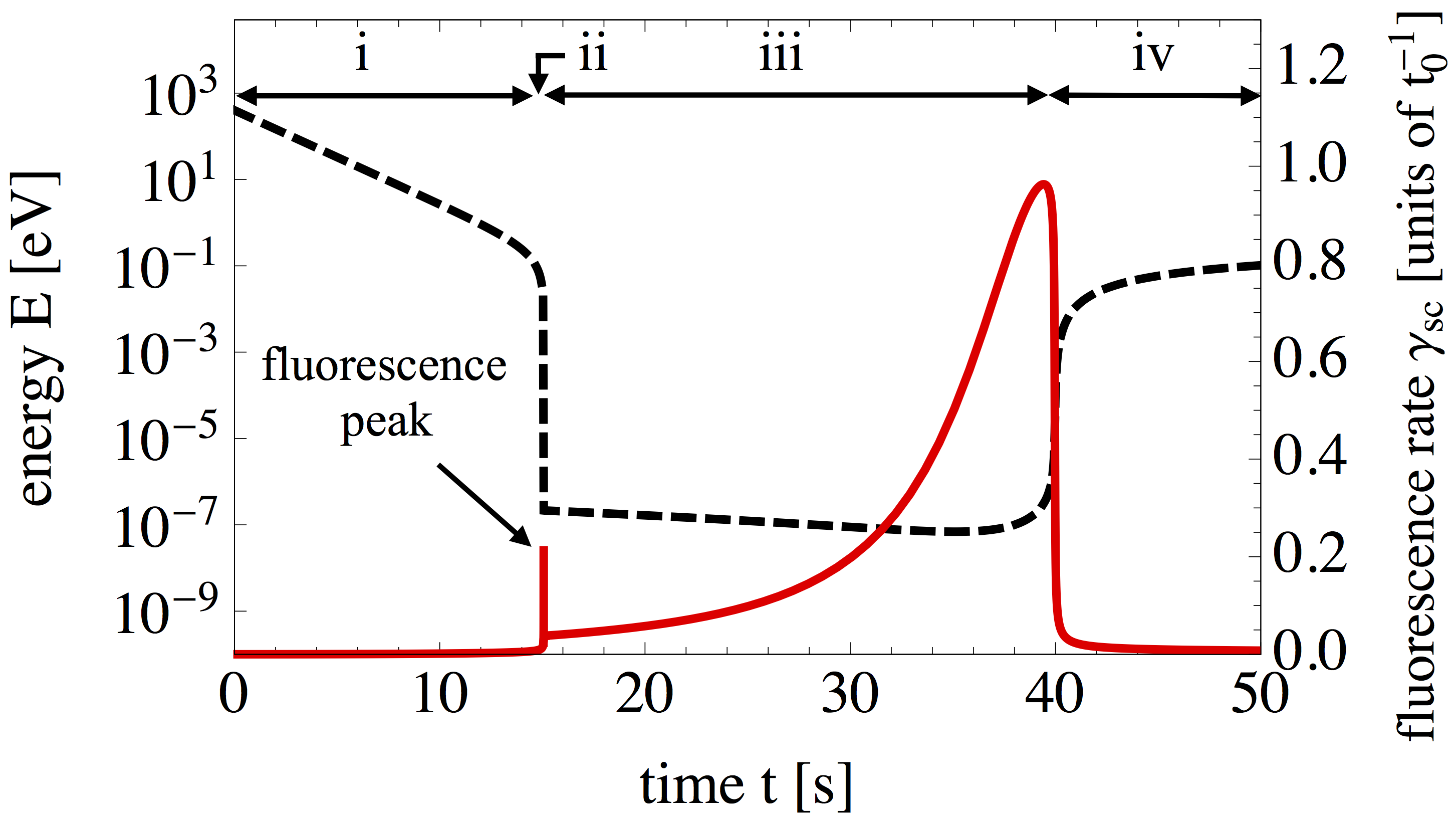}
\caption{Ion energy (black dashed curve, left axis) and scaled fluorescence rate (red curve, right axis) as a function of time during the cooling process.}
\label{Fig:1}
\end{figure}

We perform the calculation for a helium buffer gas pressure of $p=4\times 10^{-9}$\,mbar and a buffer gas temperature of 4\,K. 
The ion mobility for magnesium in a helium buffer gas is $\mu_{0}\approx 23\times 10^{-4}$\,m$^2$s$^{-1}$/V \cite{lorne}. The corresponding damping coefficient is $\gamma_{1}=0.52$/s.
The initial energy of the ion is 400\,eV and the laser parameters are $\Delta_{i}=-2\pi \times 200$\,MHz (initial detuning), $\Delta_{m}=2 \pi \times 5$\,MHz/s (scan rate), and $s_0=0.4$. The numerical results of the evaluation of Eq.\,(\ref{Eq:1}) and Eq.\,(\ref{Eq:3}) are depicted in Fig.\,\ref{Fig:1}, where the energy and the scaled fluorescence rates are shown as a function of time. 
Their behavior as a function of time can be divided into four different regimes: 
\begin{itemize}
\item[i)] the initial cooling phase is dominated by buffer gas cooling, as laser cooling is very inefficient at such high ion energies
\item[ii)] at an energy of about 1\,eV, the laser detuning corresponds roughly to the half-width of the velocity distribution, leading to a rapid cooling and reduction of the width of the velocity distribution and to a characteristic fluorescence peak in the spectra
\item[iii)] ions at the detuning-dependent Doppler cooling temperature
\item[iv)] laser heating after crossing the resonance.
\end{itemize}

The appearance of a feature such as the fluorescence peak in ii) indicates that the laser line profile and the ion velocity distribution have maximum overlap, see also \cite{rct2}. Thus, phase ii) is characterized by strong laser cooling that leads to a rapid narrowing of the initially broad velocity distribution of the ions, until a quasi-equilibrium state with a rather narrow velocity distribution is reached at the beginning of phase iii). The quasi-equilibrium is characterized by the ion temperature depending only on the laser detuning. In phase ii), the ion energy is reduced by about 5 orders of magnitude within a second. 

The temperature of an ion ensemble after the appearance of the fluorescence peak can be sufficiently low for the formation of ordered structures. Nevertheless, the observation of a fluorescence peak does not necessarily coincide with the ion cloud entering a liquid-like or crystalline state. It is neither a necessary nor a sufficient condition. Under specific experimental conditions, the ionic ensemble at that point has reached a temperature sufficient for entering a liquid-like \cite{horn} or a crystal-like state \cite{blue,new2}, as was shown in corresponding measurements in rf traps. However, we note that the characteristic `kink' in the fluorescence spectra observed in rf traps exhibits slightly more complicated dynamics, since the mechanism of rf-heating needs to be considered.

A useful quantity for the characterization of ion plasmas is the plasma parameter $\Gamma_{\text{p}}$ \cite{mal} which measures the Coulomb energy between ions relative to their thermal energy. It is defined by
\begin{equation}
\Gamma_{\text{p}} \equiv \frac{q^2}{4 \pi \epsilon_0 a_{\text{WS}} k_{\text{B}} T}
\end{equation}
where $q$ is the ion charge, $T$ is the ion temperature and $a_{\text{ws}}=(4/3\,\pi n)^{1/3}$ is the Wigner-Seitz radius \cite{kit} measuring the effective ion-ion distance at a given ion number density $n$. Ion  
\begin{figure}[h!]
		\centering
		\includegraphics[width=\columnwidth]{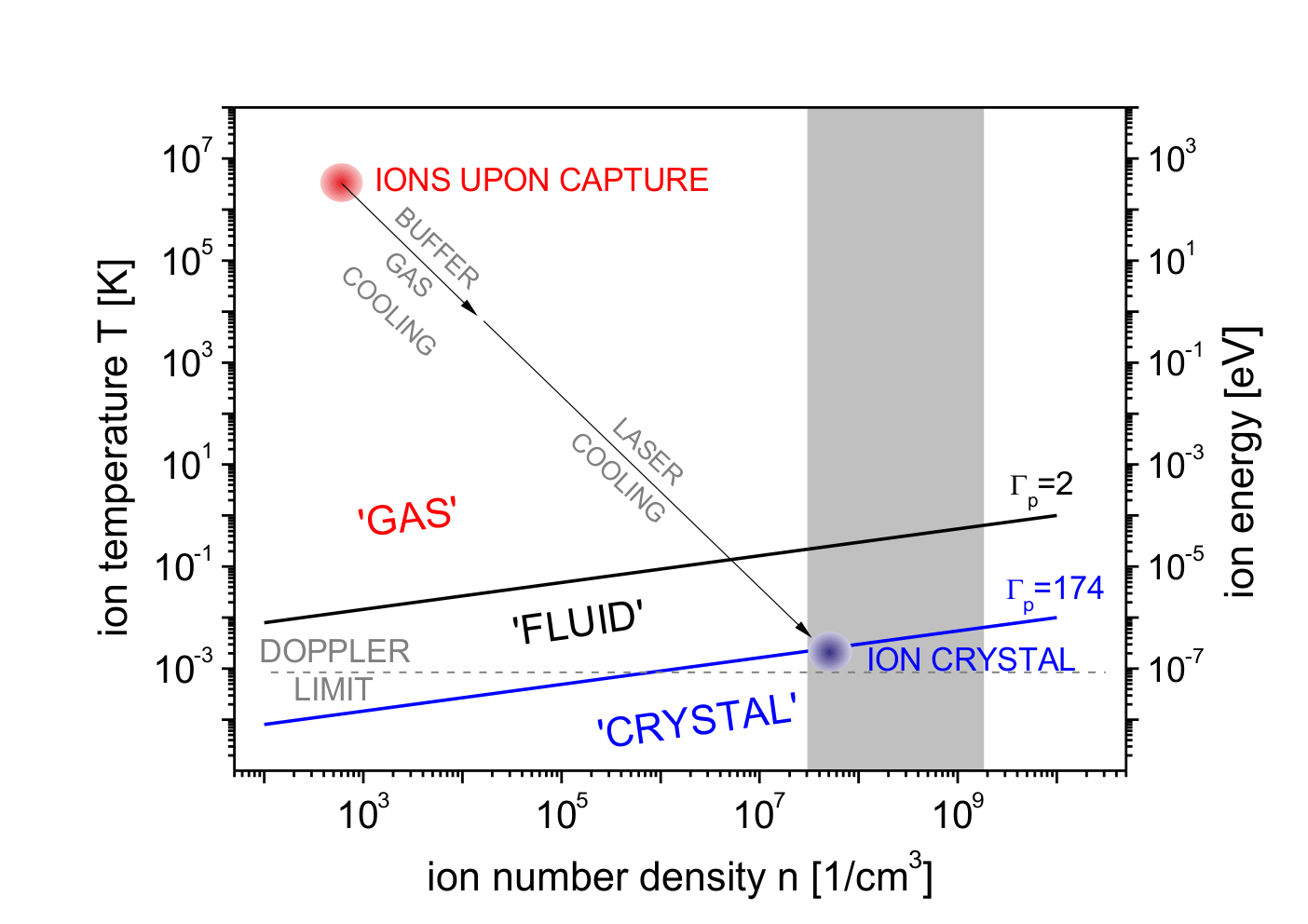}
		\caption{Phase diagram of the plasma parameter $\Gamma_{\text{p}}$ for the present conditions. Dotted line: Doppler limit of Mg$^+$ ions of 1\,mK. Gray area: possible ion number densities in thermal equilibrium. For details see text.}
		\label{gamma}
\end{figure}
cooling increases the value of $\Gamma_{\text{p}}$: commonly one speaks of a weakly correlated plasma (a gas-like state) for $\Gamma_{\text{p}} \ll 1$, and of a strongly correlated plasma for $\Gamma_{\text{p}} \gtrsim 1$. Theoretical studies predict a fluid-like behaviour for $ 174 \gtrsim \Gamma_{\text{p}} \gtrsim 2$ \cite{gil} and a crystal-like behaviour for $\Gamma_{\text{p}} \gtrsim 174$ \cite{dub2}, which has been corroborated experimentally \cite{jens}. For the magnesium ions at a density of $5 \times 10^7$/cm$^3$, this value is reached for $T \approx 5$\,mK. 

Fig. \ref{gamma} shows a phase diagram in temperature-density-space with the plasma parameter $\Gamma_{\text{p}}$ for the present trapping voltage of $U=50\,$V and the magnetic field of $B=4.1\,$T. The dotted line indicates the Doppler limit of Mg$^+$ ions of 1\,mK. The gray area shows the possible range of ion number densities $n$ in thermal equilibrium for the present trapping parameters. For sufficiently low temperature and in thermal equilibrium, the ion plasma performs a global rotation about the magnetic field axis and takes the shape of an ellipsoid of revolution with constant density \cite{dub2}. The global rotation is induced by the magnetic field used for confinement, and hence not observed in rf traps. As one important consequence, the ion number density $n$ is related to the global rotation frequency $\omega_r$ of the plasma by \cite{bre}
\begin{equation}
\label{enn}
n=\frac{2m\epsilon_0}{q^2} \omega_r (\omega_c-\omega_r),
\end{equation}
in which $\omega_r$ is bounded by the magnetron frequency $\omega_-$ and the reduced cyclotron frequency $\omega_+$ given by
\begin{equation}
\omega_{\pm} = \frac{\omega_c}{2} \pm \left( \frac{\omega_c^2}{4} - \frac{\omega_z^2}{2} \right)^{1/2}.
\end{equation}
Here, $\omega_c=qB/m$ and $\omega_z^2=qUC_2/(md^2)$ \cite{gab89}. 
For a $^{24}$Mg$^+$ ion in the present trap with a characteristic size $d=7.062$\,mm, a well-depth efficiency parameter $C_2=0.578$ and at a trapping voltage of $U=50$\,V, these frequencies are $\omega_z=2\pi\times 241.4$\,kHz, $\omega_-=2\pi\times 11.4$\,kHz and $\omega_+=2\pi \times 2.55$\,MHz.
Eq.\,(\ref{enn}) holds true as long as the Debye length
\begin{equation}
\lambda_{\text{D}}=\sqrt{\frac{\epsilon_0 k_{\text{B}} T}{n q^2}}
\end{equation}
is much smaller than the dimensions of the ion cloud \cite{bol}. For our laser-cooled Mg$^+$ ions, this is the case as $\lambda_{\text{D}}$ is of the order of $\mu$m, while the crystal size is of the order of mm.
For the possible range of $\omega_r$, Eq.\,(\ref{enn}) leads to densities between $n_{\text{min}}=n(\omega_r=\omega_-)=3.1\times 10^7$/cm$^{3}$ and $n_{\text{max}}=n(\omega_r=\omega_c/2)=1.8\times 10^9$/cm$^{3}$. 
For a given density $n$ (or, equivalently, global rotation frequency $\omega_r$), the aspect ratio $\alpha$ of the cloud (axial to radial extension) is determined by the trapping voltage $U$ according to the formalism given in \cite{bre}. 

In Fig. \ref{gamma}, the initial position of the ions in ($T,n$)-space directly upon capture into the trap is indicated (red dot), as well as the position at the end point of cooling (blue dot). Upon capture, the ions are assumed to have a density given by the measured ion number (detector F in Fig. \ref{set}) distributed over the trapping volume. After cooling, the density is determined from the measured inter-particle distance in the crystal as discussed in section \ref{densi}. Note, that the initial and final temperatures are estimated from the initial axial ion energy and the observed shell structure, respectively, the latter of which may also form below $\Gamma_{\text{p}} \approx 174$, depending on experimental detail. Note also, that the cooling path indicated serves purposes of illustration only and is not the actual (unknown) cooling path of the ions in the ($T,n$)-plane. At the end point of cooling, the ions enter an ordered state, the structure of which is the topic of the following section.

\section{Shell structure of mesoscopic ion crystals}
\begin{figure}[h!]
\begin{center}
  \includegraphics[width=0.8\columnwidth]{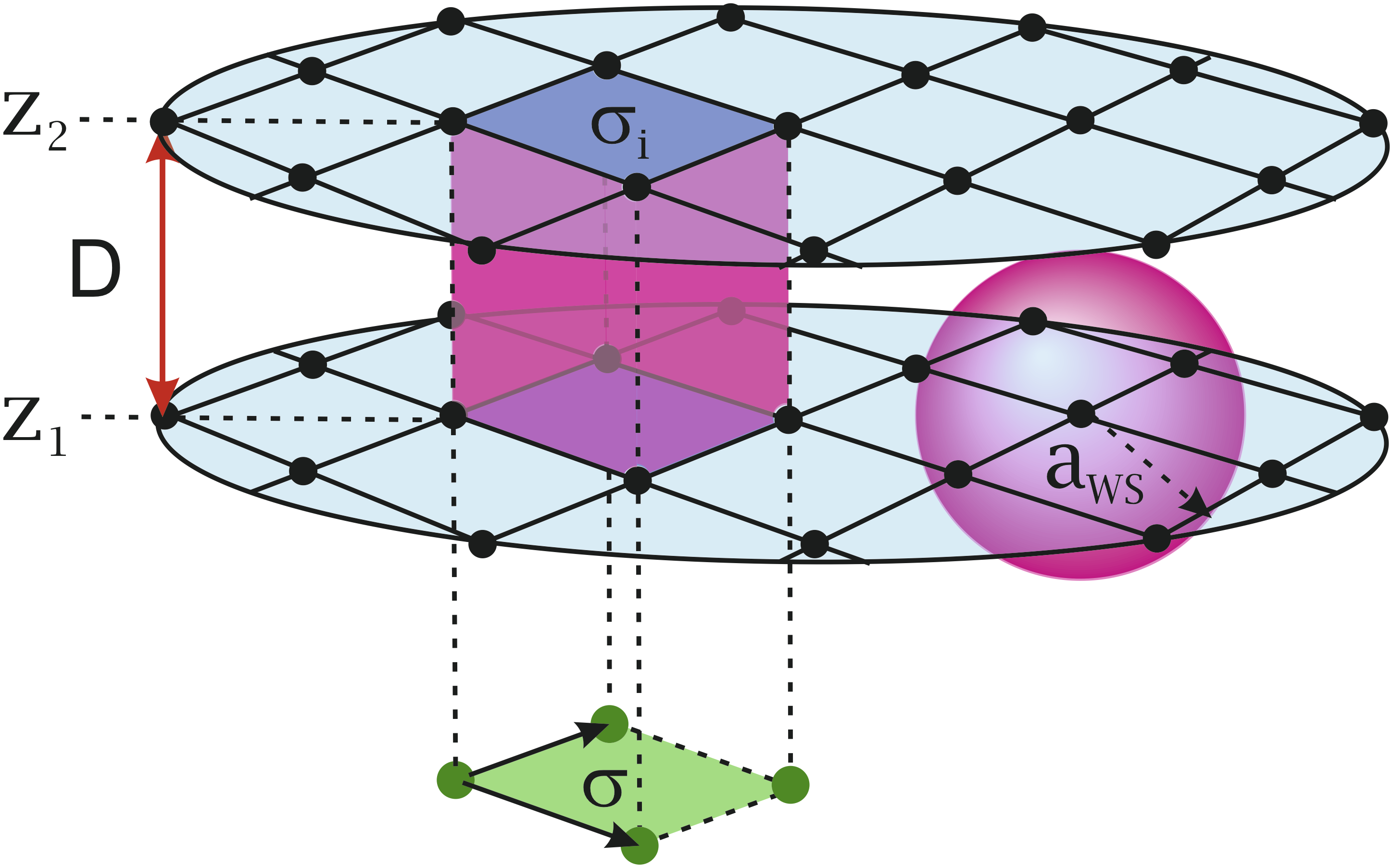}
  \caption{\small Planar-shell model geometry and the involved quantities. This model does not account for correlations in ion position between different shells.}
  \label{ebenen}
\end{center}
\end{figure}

The geometric properties of ion plasmas and the formation of ion Coulomb crystals have been described in detail for example in \cite{died,bir,dub1,dub2,dub3,mit,rich,horn,dre,dre2,bol,bol2}. For the present ion numbers
of the order of $10^3$ to $10^5$ (`mesoscopic'), and aspect ratios (axial extension to radial extension) of typically $\alpha \ll 1$, the so-called `planar-shell model' is a good approximation to describe the geometry of the confined plasmas.
It applies to the case of a spheroidal plasma with a radius sufficiently large such that the curvature of the shell planes can be neglected close to the trap axis. While in a real plasma the number of shells depends on the radial position in the crystal and decreases towards the edges, this model describes the plasma throughout as a series of $S$ parallel planes at axial positions $z_i$ with area ion number density $\sigma_i$. For the sake of simplicity, it does not explicitly account for correlations between shells, in which case one would also expect lateral offsets in the ion positions between neighbouring shells \cite{mit}.
Fig. \ref{ebenen} depicts the geometry and the involved quantities. This discussion closely follows along the lines presented in \cite{dub2}, however to the end of interpreting our measurements, it is instructive to restate some of the results given in \cite{dub2}.
For our situation, minimizing the energy per ion implies that the area charge density $q\sigma_i$ of each lattice plane is identical and that the lattice planes are spaced by a uniform distance $D$ \cite{dub2}. The total area number density $\sigma$ is the sum of all $\sigma_i$,
and the spacing $D$ is linked to $\sigma$ and $S$ via the relation
\begin{gather}
	D = \frac{\sigma}{n S}.
	\label{eq:D_layerSpacing}
\end{gather}
The contributions to the energy per particle are the self-energy of the set of $S$ planes, the energy due to the external potential, and the (negative) correlation energy associated with each 2D lattice plane. The total energy per particle reads \cite{dub2}
\begin{gather}
	\frac{E}{N}=\pi e^2\left[L\sigma-\frac{1}{6}\frac{\sigma^2}{n}\right] + \frac{U_{\text{corr}}}{N},
\end{gather}
where $2L$ is the axial extension of the crystal and $U_{\text{corr}}$ is the ion-ion correlation energy given by \cite{dub2}
\begin{gather}
	\frac{U_{\text{corr}}}{N}=\frac{e^2}{a_{\text{ws}}} \left[\frac{2\pi^2}{9}\left(\frac{\bar{\sigma}^2}{S}\right)-\frac{\eta}{2}\left(\frac{\bar{\sigma}}{S}\right)^{1/2}\right],
	\label{eq:u_corr}
\end{gather}
where $\bar{\sigma}=\sigma a^2_{\text{ws}}$ as indicated in Fig. \ref{ebenen}. In this equation, $\eta$ accounts for the Madelung energy \cite{kit} of the 2D lattice. We use the value $\eta=3.921$ of the hexagonal lattice, which has the lowest Madelung energy in 2D \cite{kit}.

A structure of $S$ parallel ion planes has a higher energy than a uniformly spread charge, which is reflected in the first term of Eq.\,(\ref{eq:u_corr}) being positive. The second term accounts for the ion-ion correlations within each plane. It is negative, which promotes the formation of a finite set of ordered planes. Hence, the number of planes that form results from the competition between these two terms.
\begin{figure}[h!]
\begin{center}
  \includegraphics[width=\columnwidth]{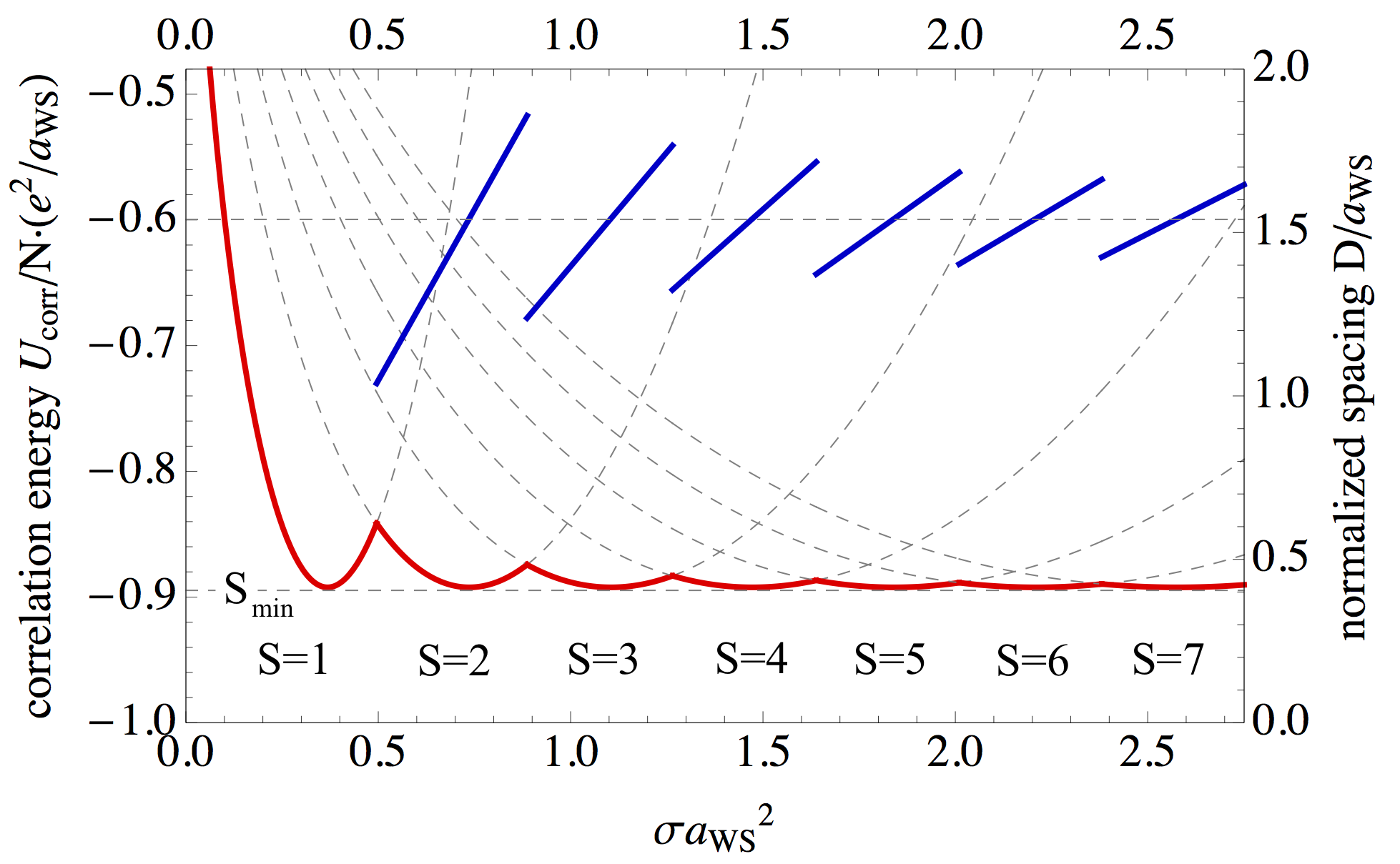}
  \caption{\small Correlation energy per particle as a function of $\bar{\sigma}=\sigma a^2_{\text{ws}}$ given for various numbers of shells $S$ (red curve, left hand scale) and the distance $D$ for which the energy per particle has a minimum as a function of the normalized area (straight blue lines, right hand scale).}
  \label{shell}
\end{center}
\end{figure}
The correlation energy per particle $U_{\text{corr}}/N$ takes a minimum value (with respect to the plane number $S$) for
\begin{equation}
S_{\text{min}}=\left[16\pi^2/(9\eta)\right]^{2/3} \bar{\sigma},
\label{eq:S_min}
\end{equation}
in which case the distance $D$ between two planes is given by
\begin{equation}
\frac{D}{a_{\text{ws}}}=\left(\frac{3\eta^2}{4\pi}\right)^{1/3}=1.54.
\label{eqd}
\end{equation}
Since the number $S$ must be an integer, $S=S_{\text{min}}$ can only be fulfilled for certain values of $\bar{\sigma}$, and the actual value of $S$ will be an integer close to $S_{\text{min}}$. 
Fig. \ref{shell} shows the correlation energy per particle $U_\text{corr}/N$ according to Eq.\,(\ref{eq:u_corr}) for shell numbers $S=1$ to $S=7$. For the given range of $\bar{\sigma}$, the shell number $S$ that gives the minimum correlation energy was chosen to calculate the inter-shell distance $D$ with Eq.\,(\ref{eq:D_layerSpacing}). 
\begin{figure}[h!]
\begin{center}
  \includegraphics[width=\columnwidth]{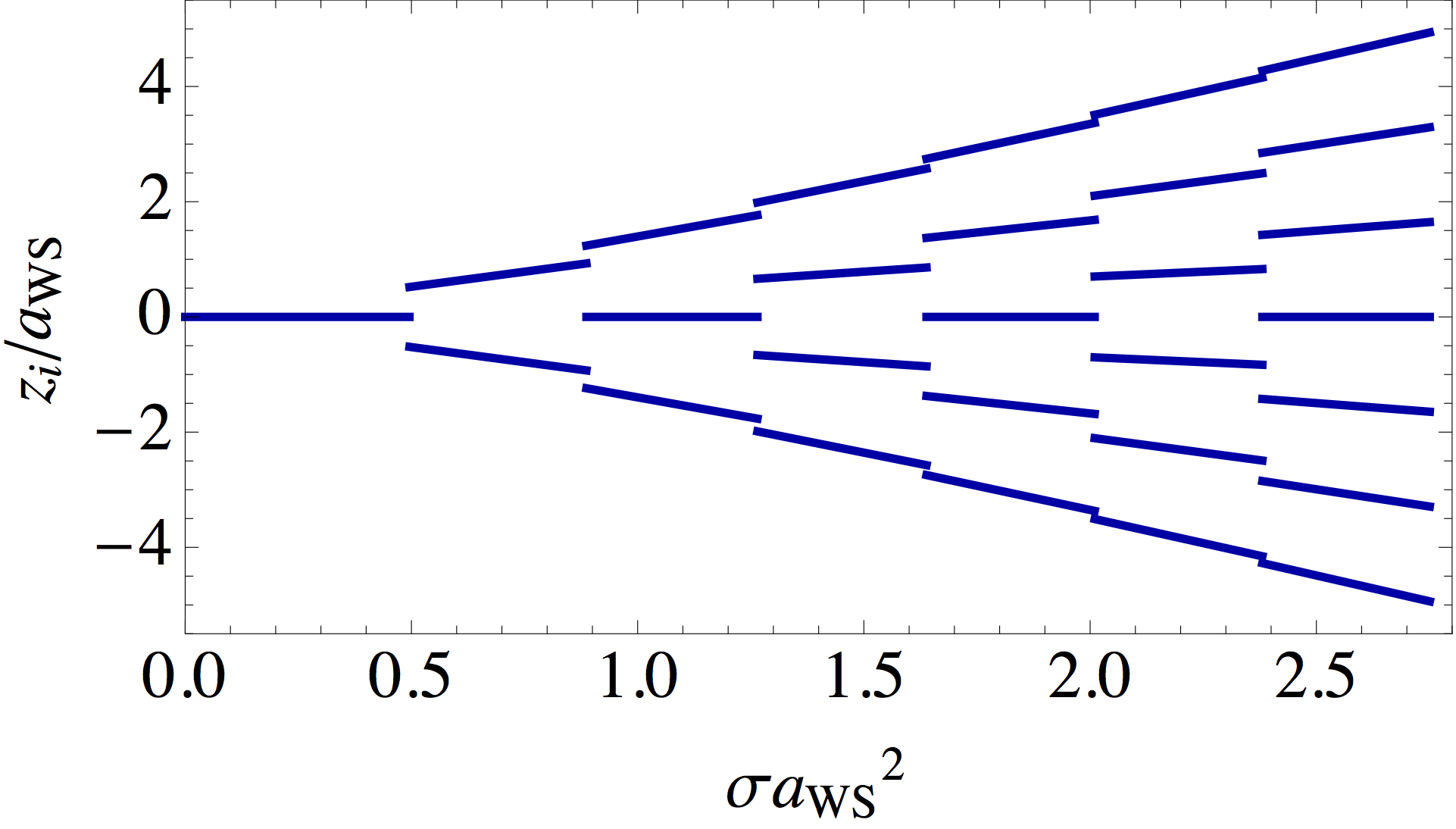}
  \caption{\small Planar-shell model prediction of the crystal shell structure as a function of $\bar{\sigma}=\sigma a^2_{\text{ws}}$.}
  \label{layer}
\end{center}
\end{figure}
One can observe that for larger $\bar{\sigma}$, the variation of the minimum correlation energy gets smaller, hence the actual correlation energy is for large shell numbers $S$ close to the minimum value. Likewise, the variation of the shell distance $D$ decreases for large shell numbers and approaches $D\approx 1.54\,a_{\text{ws}}$. The corresponding 
axial plane positions as a function of the normalized area charge density are shown in Fig. \ref{layer}.  
It illustrates the stepwise increase of the number of shells for increasing charge density, i.e. for increasing ion number.

\section{Experimental Results}
\subsection{Cooling behaviour and fluorescence}
Magnesium ions have been prepared in the trap according to the experimental cycle discussed in section \ref{setsec}. 
\begin{figure}[htb]
\begin{center}
  \includegraphics[width=\columnwidth]{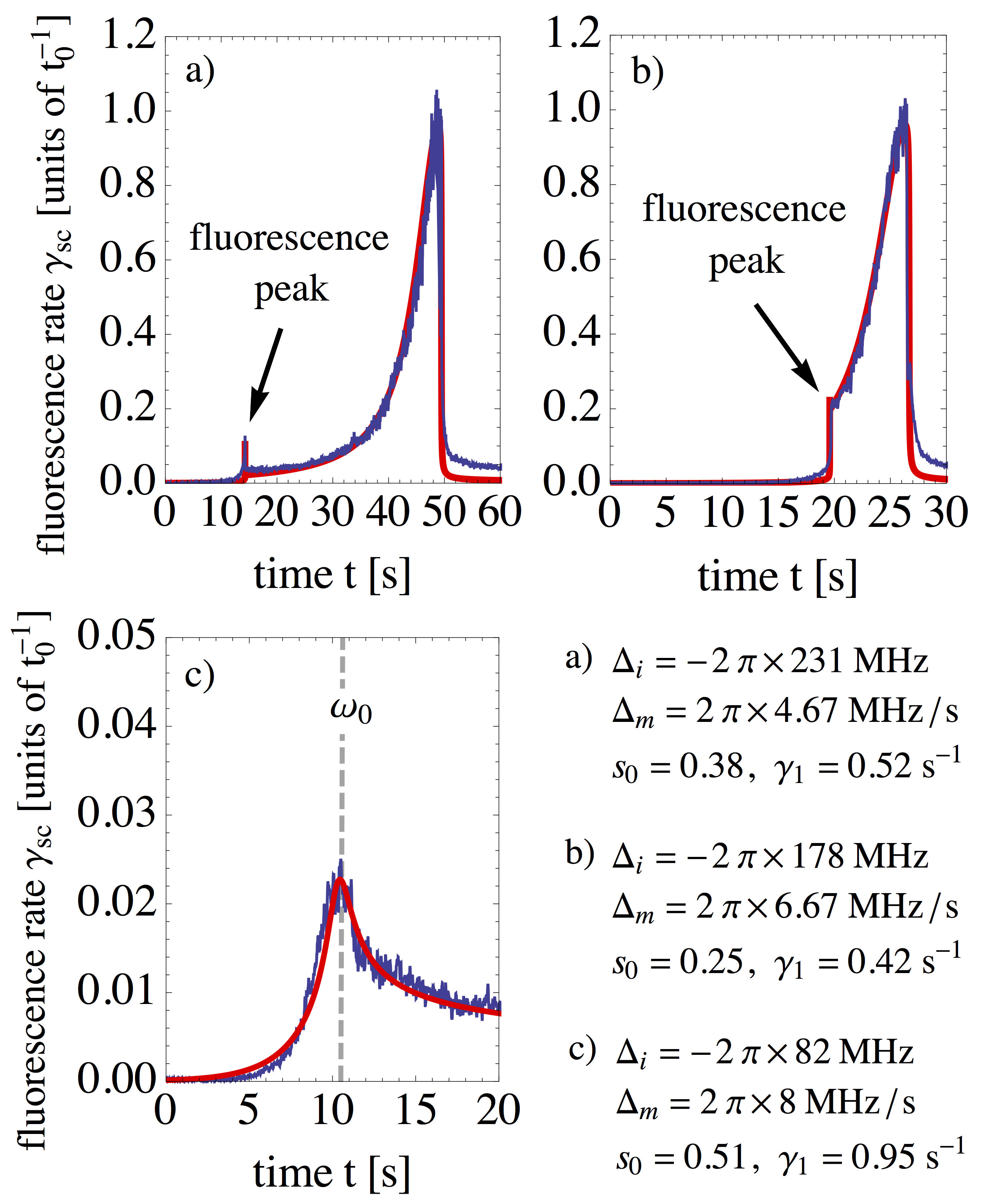}
  \caption{\small Time evolution of the fluorescence signal for three different values of the laser cooling parameters. Blue curves: data. Red curves: theory. In all cases, $E=400$\,eV and $\epsilon_1E_0=k_{\text{B}} \times 4$\,K. }
  \label{scan1}
\end{center}
\end{figure}
As indicated in section \ref{coolmod}, the cooling of the ion cloud can be monitored by observation of the fluorescence rate as a function of time. Fig. \ref{scan1} shows the time evolution of the fluorescence signal for three different sets of cooling laser parameters.
In case (a), the laser parameters were chosen such that a pronounced fluorescence peak is visible. In (b), with smaller initial detuning and higher scan rate, the influence of the laser becomes visible only close to a critical detuning of 
\begin{equation}
\Delta \lesssim -\frac{1}{\sqrt{3}} \frac{\Gamma}{2} \sqrt{1+s_0},
\end{equation}
such that the fluorescence peak is not pronounced. In (c), the scan rate is so high that phase iii (laser cooling equilibrium) is never reached, and the ions remain at a temperature of around 100\,K before phase iv (laser heating) sets in. The red curves in Fig. \ref{scan1} show the predictions of the model presented in section \ref{coolmod} according to Eq.\,(\ref{Eq:3}), where the value of the parameter $\gamma_1$ has been adjusted in each case in order to apply the single-particle model to the experimental many-particle system. In all cases discussed in this work, the value of $\gamma_1$ lies within a factor of 2.5 which is on account of a variation of the helium gas pressure between different experimental runs.

\begin{figure}[h!]
\begin{center}
  \includegraphics[width=\columnwidth]{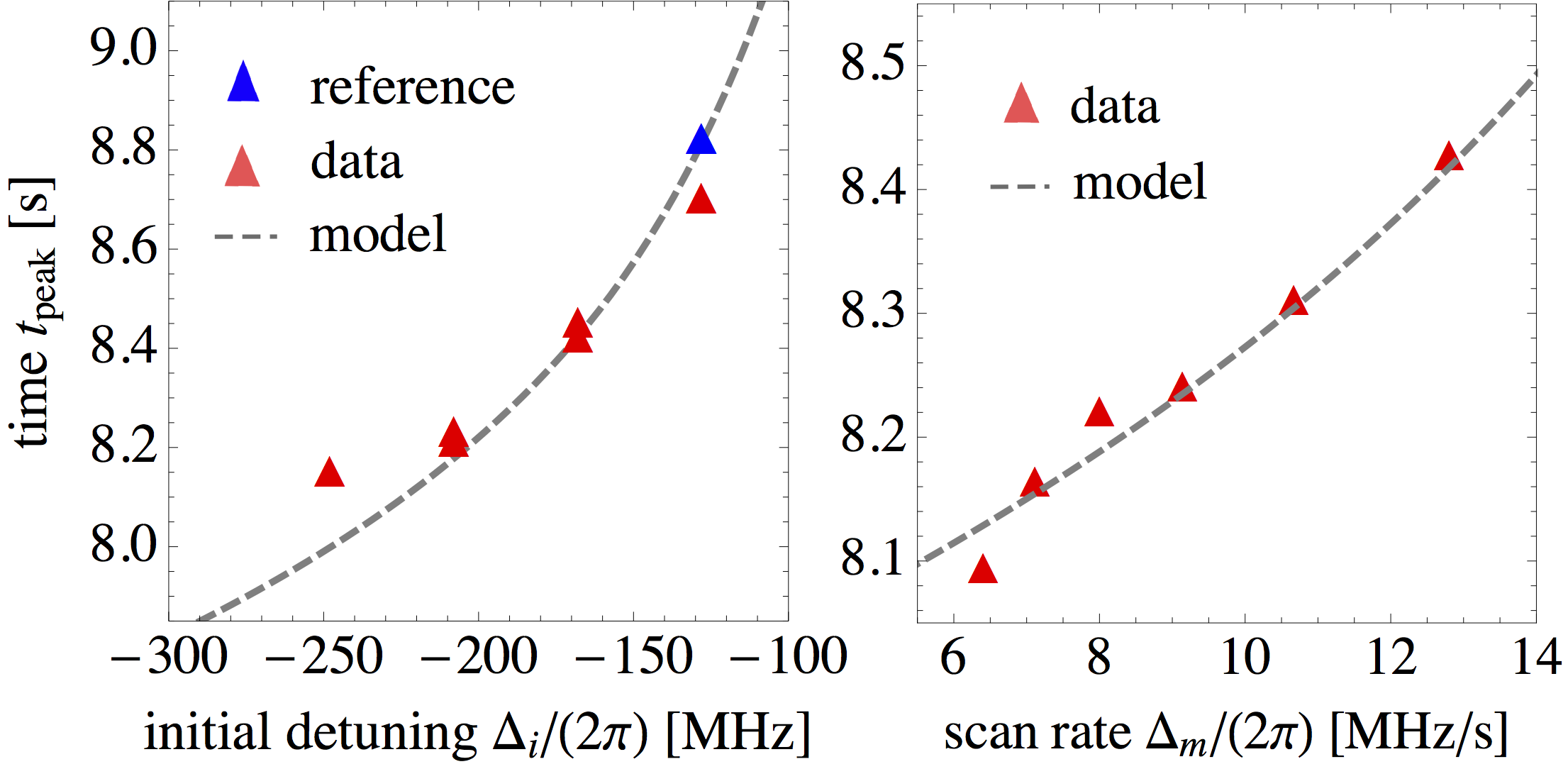}
  \caption{\small Left: Appearance time of the fluorescence peak as a function of the initial laser detuning for a constant scan rate of $\Delta_m=2\pi \times 8$\,MHz/s. Right: same as a function of the laser scan rate for constant initial detuning $\Delta_i=-2\pi\times 208$\,MHz.}
  \label{scan2}
\end{center}
\end{figure}

We have performed systematic measurements of the appearance time $t_{\text{peak}}$ of the fluorescence peak as a function of laser parameters. Fig. \ref{scan2} shows the appearance time as a function of the initial laser detuning and the laser scan rate, respectively. 
To calibrate the measurements, we have done an independent measurement of the damping constant $\gamma_1$ as input parameter for the model presented in section \ref{coolmod}, assuming again an initial ion energy of 400\,eV. The model prediction for the peak appearance time is plotted in Fig. \ref{scan2} as dashed lines. The left hand graph shows the appearance time as a function of the initial detuning for constant scan rate. The right hand graph shows the same as a function of the scan rate for constant initial detuning.
Obviously, the appearance time increases with decreasing initial detuning $|\Delta_i|$, since the energy taken away per cooling cycle decreases. Also, the appearance time increases with increasing scan rate $\Delta_m$, which can  be understood in that the laser spends less time at large detuning where the dissipated energy per cooling cycle is highest. The cooling model from section \ref{coolmod} (dashed lines) agrees well with the data.

\subsection{Crystal formation}
When the laser cooling reduces the kinetic energy of the confined ions sufficiently, they `freeze' in the effective potential given by the trap and their mutual Coulomb interactions, as has been demonstrated in numerous experiments, see for example \cite{died,bir,dre,dre2,horn,mit,rich}.  
\begin{figure}[h!]
\begin{center}
  \includegraphics[width=\columnwidth]{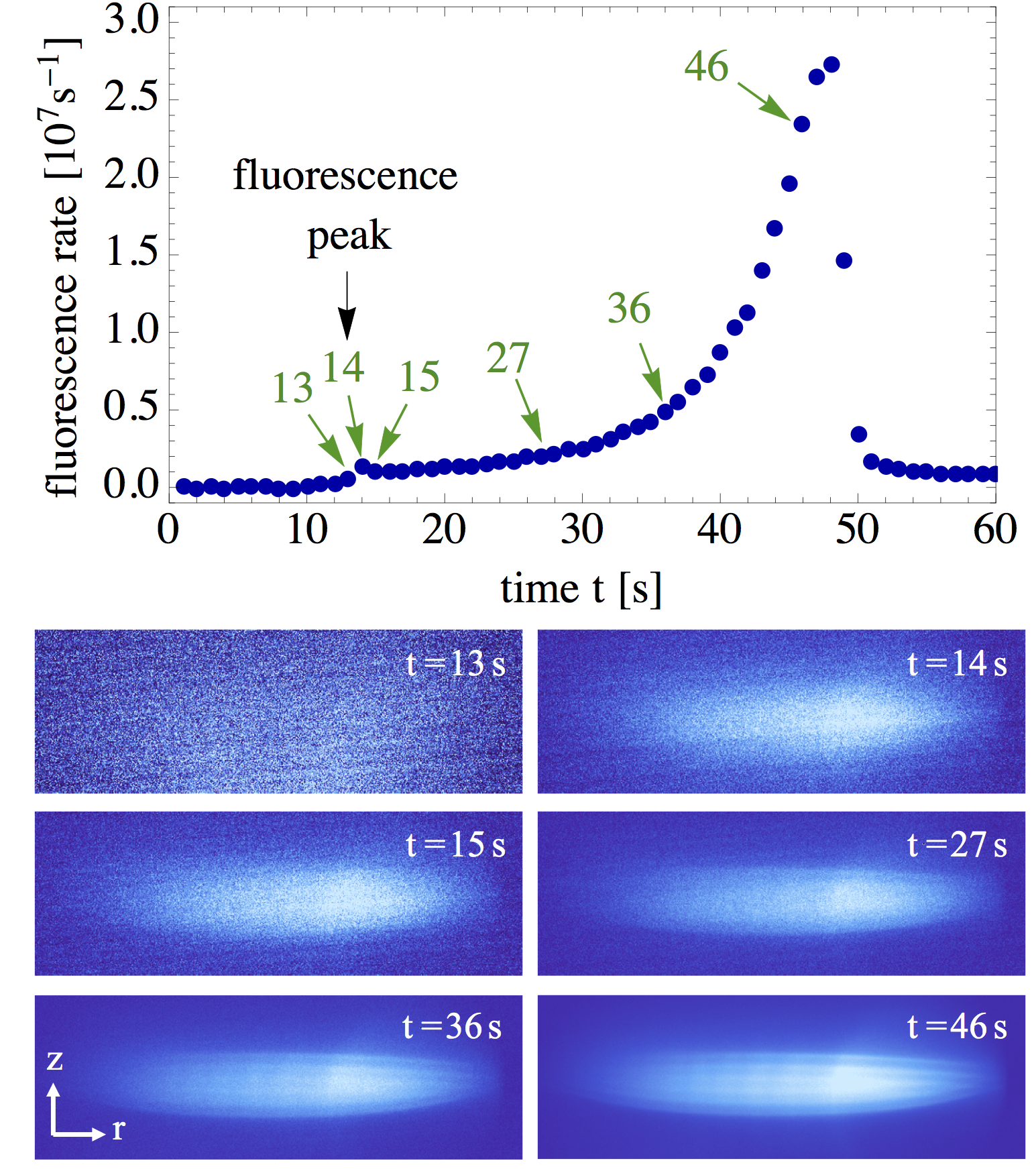}
  \caption{\small Top: Detected fluorescence during ion cooling and crystal formation extracted from fluorescence images. Bottom: Selected images of the ion cloud during cooling. All images shown are false-colour images of the UV fluorescence with light colour representing the highest intensity.}
  \label{res1}
\end{center}
\end{figure}
Unlike crystals known from solid-state physics, these Coulomb crystals have no intrinsic binding force, but a mutual repulsion inside the common external potential well of the trap.
In a Penning trap, the equilibrium state of an ion crystal is an ordered structure as described in \cite{dub2} that performs a global rotation about the trap's central axis ($z$-axis) at a frequency $\omega_r$ set by the initial conditions, and bounded by the magnetron frequency $\omega_-$ and the reduced cyclotron frequency $\omega_+$ \cite{bol}, as discussed in section \ref{coolmod}. 
The global rotation at $\omega_r$ leads to a smearing-out of the observed structure in the $x$- and $y$-directions, if the exposure time is not negligible with respect to the inverse of the rotation frequency.  
The visibility of the shells, however, is to a large extent unaffected by this, such that the present images resolve the shell structures even for long exposure times. 

For the conditions present in our experiment, the global rotation frequency has been determined to be close to the magnetron frequency, and hence the clouds have aspect ratios much smaller than unity, i.e. they are of oblate shape.

Fig. \ref{res1} shows the measured fluorescence rate 
as a function of time during ion cooling, and shows ion images at selected times. This data is from the same measurement as the data in Fig. \ref{scan1} (a).
As expected, the initially diffuse ion cloud increases in density during cooling. In particular, when crossing the fluorescence peak at $t=14$\,s, there is a sudden increase in density from the diffuse situation at $t=13$\,s to the denser distributions at $t=14$\,s and $t=15$\,s. The shell structure becomes visible a few seconds after that fluorescence peak, from about $t=27$\,s on, and intensifies as the laser is further scanned towards resonance, see $t=36$\,s and $t=46$\,s. Note, that there is no indication of any crystalline feature in the images when crossing the fluorescence peak between $t=13$\,s and $t=15$\,s. 

\subsection{Geometric structure}
\label{densi}
The ion crystals under investigation consist of several thousands of Mg$^+$ ions. Hence, they fall into the category of `mesoscopic' ion crystals which are large  
\begin{figure}[h!]
\begin{center}
  \includegraphics[width=\columnwidth]{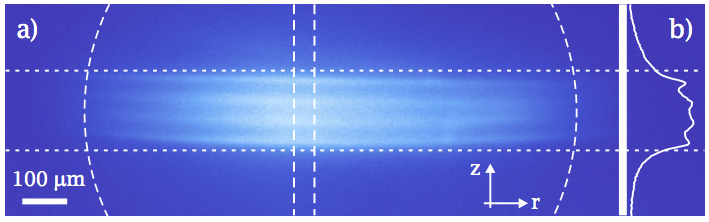}
  \caption{\small a) CCD image of an ion crystal with scale given. The circle indicates the trapping region visible to the camera. b) Cross section of the fluorescence, showing 4 shells.}
  \label{pic}
\end{center}
\end{figure}
enough to display a shell structure and are still subject to surface effects, not having reached the universal lattice structure of macroscopic crystals \cite{dub2}. Fig. \ref{pic} (a) shows a detailed CCD image of a crystal. 
\begin{figure}[h!]
	\centering
	\includegraphics[width=1\columnwidth]{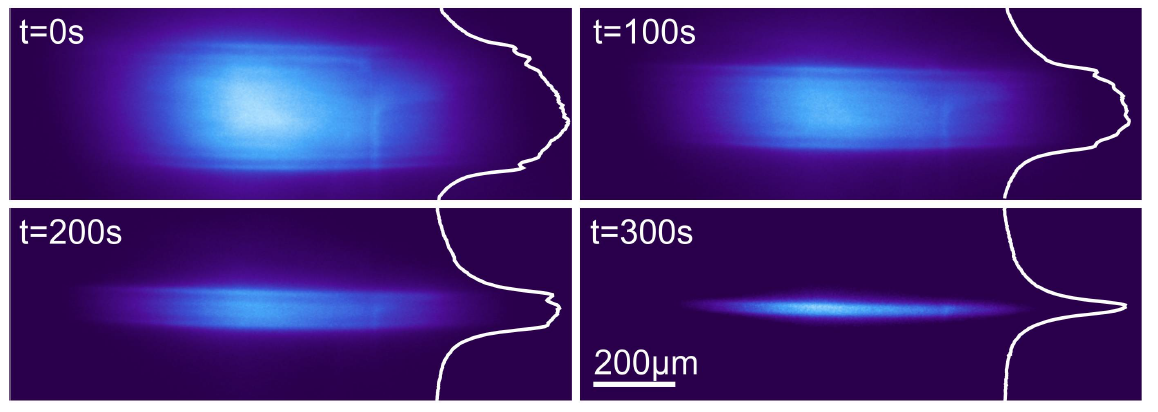}
	\caption{Images of a Mg$^+$ ion crystal with the number of ions and hence shells decreasing with time. For details see text.}
	\label{fourclouds}
\end{figure}
The figure also indicates the observable part of the trapping region as the interior of the dashed circle, and the vertical section in the middle from which the cross section (b) is taken. In this way, all comparable images in this work have been taken and evaluated.
Four images of a mesoscopic crystal and corresponding cross sections studied for 7 minutes are shown in Fig. \ref{fourclouds}, stored in a magnetic field of $B=4.1$\,T and a trapping voltage of $U=50$\,V.
The structure with parallel planar shells is visible both in the images and the cross sections. The presented images have a temporal separation of 100\,seconds and an exposure time of 5\,seconds each. 
The buffer gas pressure is such that the number of ions and the number of lattice planes decreases on this timescale, allowing a convenient observation of structures with varying ion number. For each image, the contrast has been adjusted individually such that the shell structure is clearly visible. To show the real intensity relations, the intensity profile through the crystal center is also displayed. Here, each shell appears as a small deviation from the average crystal profile.
\begin{figure*}
	\includegraphics{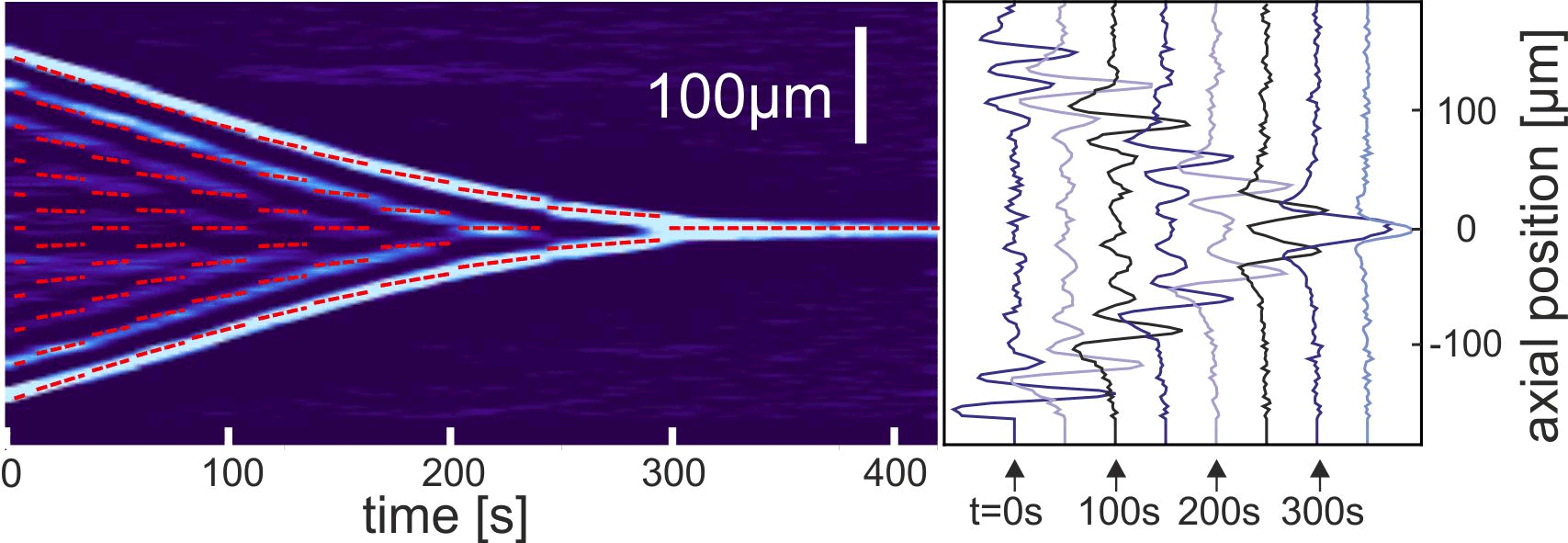}
	\caption{Left: temporal evolution of the ion crystal cross section depicted in Fig. \ref{fourclouds}. Red dots are theory values from the planar-shell model, for details see text. Right: the corresponding residual cross sections. Up to eleven crystal shells can be seen.}
	\label{eiffel}
\end{figure*}

The temporal evolution of this shell structure is shown in Fig. \ref{eiffel} (left). It displays the measured cross sections as indicated in Fig. \ref{pic} as a function of time. Each cross section is integrated over 5 seconds exposure time, such that for the 420 seconds, the 84 displayed cross sections are obtained.
A number of selected cross sections are shown in the right part of Fig. \ref{eiffel} as residual cross sections, in which the measured cross section is normalized to its running average. For large numbers of shells, the contrast in the crystal center is small and for the largest crystal (leftmost cross section), the individual shells are not clearly visible. The total number of shells can, however, be determined by comparison with smaller crystals because the shell positions and shell distances are preserved.

In Fig. \ref{eiffel}, the parameter $\bar{\sigma}$ was calculated from the integrated fluorescence of a cross section. With this method the area density was determined for each frame, and the vertical shell positions $z_i$ were calculated according to Eq.\,(\ref{eq:D_layerSpacing}). 
The calculated positions $z_i$ of the crystal shells are shown by the red lines in Fig. \ref{eiffel}. In addition to the number of shells $S$ and the shell positions, also the times - and corresponding densities - where transitions from $S+1$ to $S$ occur are predicted correctly.

From the results shown in Figs. \ref{fourclouds} and \ref{eiffel}, the value of the Wigner-Seitz radius was determined as $a_{\text{ws}}=19.1\,\mu$m, corresponding to a density of $n=3.4\times 10^7\,$/cm$^{3}$, close to the minimum density of $3.1\times 10^7$/cm$^{3}$. The corresponding rotation frequency and aspect ratio are $\omega_r=2\pi\times 12.2$\,kHz and $\alpha \approx 1/24$.
The total number of particles stored in the trap is $N= 4/3\, \pi z_0 r_0^2 \times n$, where $z_0$ and $r_0=z_0/\alpha$ denote the axial and the radial crystal radius, respectively. For the time $t=0$\,s one finds $z_0=150\,\mu$m and thus $r_0=3.6\,$mm, which gives a total particle number of $N\approx 3\times 10^5$. Assuming that the visible volume is determined by the laser beam with a waist of $w_0=1\,$mm, about $3\times 10^4$ Mg$^+$ ions are visible in the images.

\subsection{Two-Species Ion Crystals}
Two-species ion crystals \cite{new1,new2} were formed by sympathetic cooling of dark ions after their injection into a cloud of laser-cooled magnesium ions. 
Two-species crystals composed of Mg$^+$ and ions with masses $m$=2\,u (H$^+_2$), $m$=12\,u (C$^+$), $m$=28\,u (N$^+_2$), and $m$=44\,u (CO$^+_2$) were studied, such that a mass-to-charge ratio range of 2 to 44 was covered.
The characteristics of the mixed-ion crystals can be classified with regard to the mass-to-charge ratio of the sympathetically cooled species. After injection of CO$^+_2$ into the trap, the axial extent of the ion cloud increased due to the larger total number of ions, since the density remained unchanged. Consequently, additional crystal shells were formed. 

\begin{figure*}
	\includegraphics{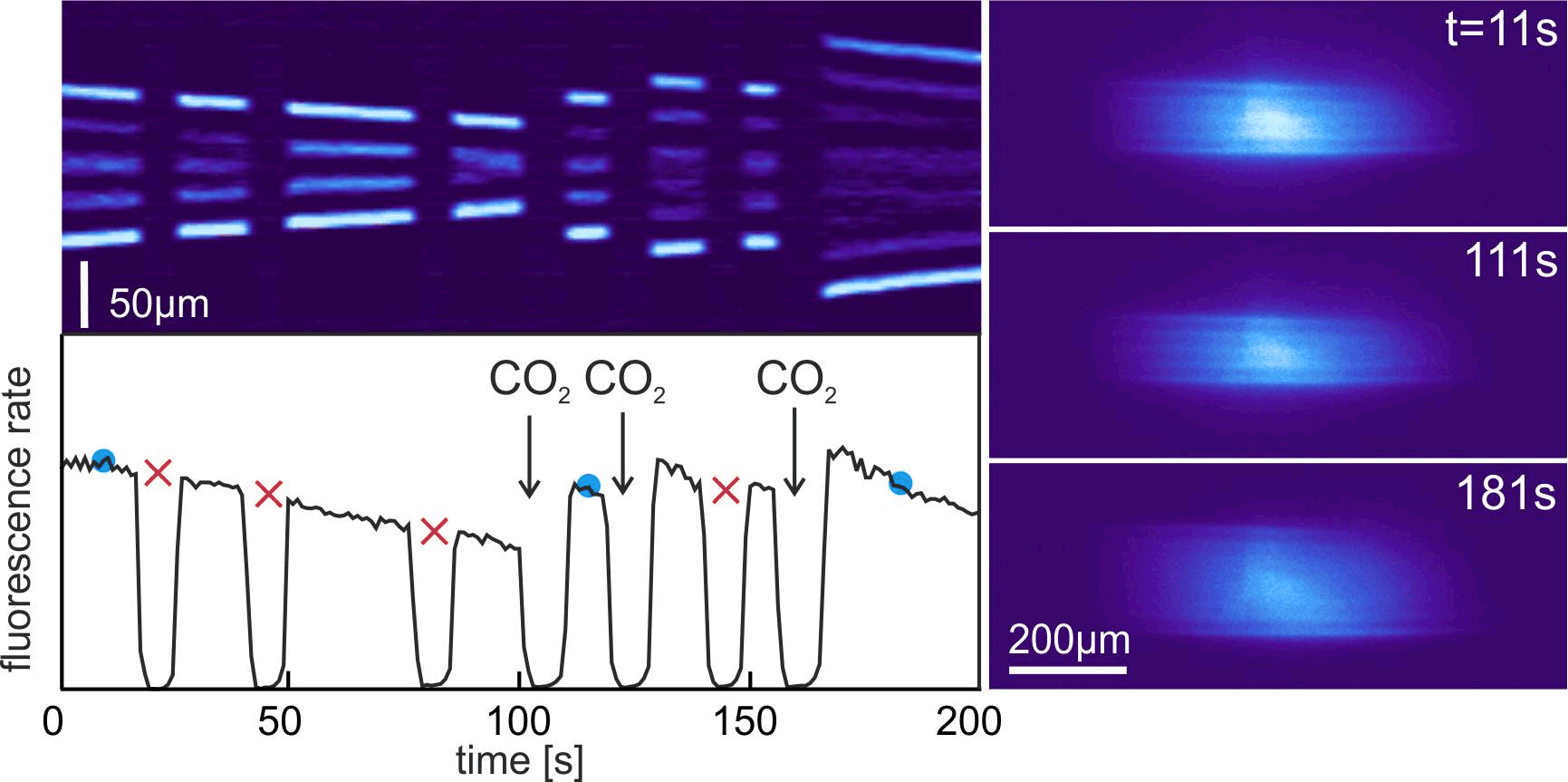}
	\caption{Fluorescence signal (bottom left) of a Mg$^+$ cloud during the consecutive injection of three bunches of CO$_2^+$ ions ($m$=44\,u) and four capture switching cycles without ion loading (red `X'). Three crystal images (right) show the crystal structure at different times (blue dots in fluorescence signal). The temporal evolution of the cross section for each video frame (top left) reveals the temporal evolution of the crystal structure.}
	\label{fig:m44_sympathetic}
\end{figure*}

Fig. \ref{fig:m44_sympathetic} shows an example of the temporal fluorescence evolution, cloud images before and after loading of CO$_2^+$, and the evolution of the crystal structures. Dips in the fluorescence signal are produced by switching of the capture electrode during ion injection. Red crosses mark the injection processes, for which the parameters were intentionally chosen such that no CO$_2^+$ was captured into the trap. 
After the dip, the fluorescence signal recovers to the value expected without switching of the capture electrodes.

The injection of CO$^+_2$ becomes apparent by an increase of the fluorescence signal and the number of crystal shells.
Since the CO$_2^+$ ions are not fluorescing, it is not possible to determine their distribution directly. However, one may assume that CO$_2^+$ and Mg$^+$ are radially separated, since upon loading of CO$_2^+$ ions, the overall fluorescence increases while the fluorescence per shell remains approximately constant. This indicates that the number of Mg$^+$ in the observed volume increased, because the CO$_2^+$ ions accumulate at larger, unobservable radii and push the Mg$^+$ ions to the trap center and thus into the laser beam. Also, the clearly observable shell structure suggests a temperature below 100\,mK, at which Mg$^+$ and CO$_2^+$ ions should undergo centrifugal separation, similar to the cases discussed in \cite{sep1,sep2}.

After injection of N$^+_2$ ions ($m$=28\,u), effects similar to the case of CO$^+_2$ were observed. For injected C$^+$ ions ($m$=12\,u), the crystal structure was conserved and the axial cloud extent increased due to the formation of additional crystal shells. However, unlike the cases of N$_2^+$ and CO$_2^+$, the injection of C$^+$ caused no increase of the total fluorescence. Since the cloud extent increased nonetheless, this means that the fluorescence per shell was reduced. The fluorescence reduction cannot be explained by significant heating since the crystal structure was preserved, which corroborates a radial separation of the ion species. In this case, the lighter C$^+$ ($m$=12\,u) ion accumulate in the trap center, whereas the Mg$^+$ ions are forced to larger radii from which they contribute less to the detected fluorescence.

After injection of H$_2^+$ - the lightest ion species under investigation - a loss of fluorescence per volume was observed. Here, the fluorescence signal decreased to a small fraction of the initial value after the capture of H$_2^+$, whereas the axial cloud extent remained constant. In analogy to the discussion of the other ion species, this indicates a radial separation of H$_2^+$ and Mg$^+$ with the lighter hydrogen in the center of the trap. However, no crystalline shell structure could be observed in this two-species ion cloud, either since the signal-to-noise ratio was insufficient or because the ordered structure was actually lost. As a consequence, a fluorescence decrease due to an unidentified heating process cannot be entirely excluded in this case. Nevertheless, the assumption of centrifugal separation is justified since the mass difference between Mg$^+$ and H$^+_2$ is the largest of all ion species under investigation. Apparently, centrifugal separation is to be expected even for comparatively large temperatures.

Overall, these investigations prove that two-species ion crystals were formed over a large range of charge-to-mass ratios of the involved species with large numbers of dark ions. Although centrifugal separation is a possible limitation for spectroscopy of sympathetically cooled species, this shows that sympathetic cooling down to crystalline structures is possible also for externally produced ions at initially high energies. This concept can be extended to multi-species crystals with ions from different sources. 

\section{Summary}
We have applied a combination of buffer gas cooling and laser cooling to externally produced Mg$^+$ ions captured and confined in a Penning trap. This technique has been found to reduce the ion kinetic energy by eight orders of magnitude within seconds, leading to the ions entering a crystalline state. We have observed the temporal evolution of the ion fluorescence that reflects the ion kinetic energy and find agreement with a model of combined buffer gas and laser cooling. We have studied the geometric properties of the resulting ion crystals and find agreement with the planar-shell model which applies to ion crystals of the present size, i.e. so-called `mesoscopic' ion crystals consisting of several thousands to several tens of thousands of ions. 

When other ion species are captured and confined together with already stored and cooled Mg$^+$ ions, they are sympathetically cooled and together form two-species ion crystals, with properties depending on the combination of mass-to-charge ratios, in agreement with theory of centrifugal separation. The present findings demonstrate highly efficient cooling of ions in a Penning trap upon capture from external sources at medium to high transport energies, including sympathetic cooling of ion species for which no laser-cooling transition exists. This facilitates precision spectroscopy of confined ions from external sources as it allows efficient cooling and hence a suppression of the influence of the Doppler effect. When the initial buffer gas cooling is spatially or temporally separated from the laser cooling, this method is also suitable for sympathetic cooling of highly charged ions into a Doppler-free regime.

\section{Acknowledgement} We thank Hamamatsu for loan of the CCD camera, and Zoran Andelkovic, Bernhard Maa\ss, Alexander Martin, Oliver Kaleja, Kristian K\"onig, J\"org Kr\"amer, Tim Ratajczyk and Rodolfo Sanchez for their support in the conduction of the experiment. We gratefully acknowledge the
support by the Federal Ministry of Education and Research
(BMBF, Contract Nos. 05P15RDFAA, the Helmholtz International Centre for FAIR
(HIC for FAIR) within the LOEWE program by the federal state
Hessen, the Deutsche Forschungsgemeinschaft (DFG contract BI 647/5-1) and
the Engineering and Physical Sciences Research Council
(EPSRC). S.S. and T.M. acknowledge support
from HGS–HIRe. The experiments have been performed
within the framework of the HITRAP facility at the Helmholtz
Center for Heavy Ion Research (GSI) at Darmstadt and the
Facility for Antiproton and Ion Research (FAIR) at Darmstadt.

\end{document}